\documentclass[runningheads]{llncs}

\usepackage[review,year=2024,ID=3]{cv4dc}
\usepackage{cv4dcabbrv}

\usepackage{graphicx}
\usepackage{booktabs}
\usepackage{array}
\usepackage[accsupp]{axessibility}  % Improves PDF readability for those with disabilities.

\usepackage[pagebackref,breaklinks,colorlinks,citecolor=cvdcblue]{hyperref}
\usepackage{orcidlink}

\begin{document}

\title{SkinMamba: A Precision Skin Lesion Segmentation Architecture with Cross-Scale Global State Modeling and Frequency Boundary Guidance} 

\titlerunning{Abbreviated paper title}

% TODO REVIEW: Replace with your author list. 
% \author{Shun Zou\inst{1} \orcidlink{0000-1111-2222-3333}\and
% MingyaZhang\inst{2}$^{\ast}$\orcidlink{1111-2222-3333-4444} \and
% Yi Zou\inst{3}\orcidlink{2222--3333-4444-5555}
% \and
% Yi Zou\inst{3}\orcidlink{2222--3333-4444-5555}
% \and
% Xiuguo Zou\inst{1}\orcidlink{2222--3333-4444-5555}
% }
\author{
Shun Zou\inst{1}\textsuperscript{*}\orcidlink{0009-0005-6311-6718} 
\and
Mingya Zhang\inst{2}\textsuperscript{*}\orcidlink{0009-0007-2233-2689} \and
Bingjian Fan\inst{1}\orcidlink{0009-0005-9859-9369} 
\and
Zhengyi Zhou\inst{1}\orcidlink{0009-0004-9246-7202}
\and
Xiuguo Zou\inst{1}\textsuperscript{†}\orcidlink{0000-0002-8074-7555}
}
\footnotetext[1]{\textsuperscript{*} These authors contributed equally to this work.}
\footnotetext[2]{\textsuperscript{†} Corresponding author.}

\authorrunning{F.~Author et al.}

\institute{College of Artificial Intelligence, Nanjing Agricultural University
\email{\{zs,fbj,19222120\}@stu.njau.edu.cn}, \email{zouxiuguo@njau.edu.cn}
\and
State Key Laboratory for Novel Software Technology, Nanjing University
\email{dg20330034@smail.nju.edu.cn}\\
}
\maketitle

\begin{abstract}
Skin lesion segmentation is a crucial method for identifying early skin cancer. In recent years, both convolutional neural network (CNN) and Transformer-based methods have been widely applied. Moreover, combining CNN and Transformer effectively integrates global and local relationships, but remains limited by the quadratic complexity of Transformer. To address this, we propose a hybrid architecture based on Mamba and CNN, called SkinMamba. It maintains linear complexity while offering powerful long-range dependency modeling and local feature extraction capabilities. Specifically, we introduce the Scale Residual State Space Block (SRSSB), which captures global contextual relationships and cross-scale information exchange at a macro level, enabling expert communication in a global state. This effectively addresses challenges in skin lesion segmentation related to varying lesion sizes and inconspicuous target areas. Additionally, to mitigate boundary blurring and information loss during model downsampling, we introduce the Frequency Boundary Guided Module (FBGM), providing sufficient boundary priors to guide precise boundary segmentation, while also using the retained information to assist the decoder in the decoding process. Finally, we conducted comparative and ablation experiments on two public lesion segmentation datasets (ISIC2017 and ISIC2018), and the results demonstrate the strong competitiveness of SkinMamba in skin lesion segmentation tasks. The code is available at \href{https://github.com/zs1314/SkinMamba}{https://github.com/zs1314/SkinMamba}.
\end{abstract}

\section{Introduction}
\label{sec:intro}
According to the 2023 global cancer statistics \cite{siegel2023cancer}, tens of thousands of people die annually from malignant skin lesions, with melanoma, a highly lethal skin cancer, becoming one of the fastest-growing cancers worldwide. In recent years, the advancement of computer technology has led to the widespread use of computer-aided diagnosis (CAD) in the medical field \cite{chan2020computer,tang2020novel,leming2023challenges,hadjiiski2023aapm}. Automated skin lesion segmentation systems assist medical professionals in rapidly and accurately identifying lesion areas. In medical image segmentation, we mainly use Convolutional Neural Networks (CNN) \cite{azad2024medical} and Vision Transformers (ViT) \cite{xiao2023transformers,zhou2023nnformer,wu2024medsegdiff}. However, both have certain limitations. CNN excels at capturing local features but fails to fully extract global features, which reduces segmentation accuracy. ViT captures long-range dependencies, enabling efficient global modeling and focusing on key areas of the image, but its quadratic complexity results in a high computational burden. Although some studies have attempted to explore efficient attention mechanisms \cite{Touvron2020TrainingDI,9711179,tu2022maxvit,liu2021Swin} or combine CNNs with ViTs to create lightweight hybrid models \cite{rahman2023medical,yuan2023effective,Lin2023ScaleAwareMM}, these approaches often come at the cost of reducing the ability to capture global features. The trade-off between efficient global modeling and computational efficiency remains unresolved.

In recent years, State Space Models (SSM) \cite{kalman1960new, fu2023hungry,gupta2022diagonal}  have attracted the attention of many researchers. S4 \cite{gu2022efficiently} initially applied SSM to the deep learning field, but its performance still lagged behind CNN and Transformer. Thanks to Mamba \cite{mamba2}, which enhances SSM through an efficient selective scanning mechanism, it not only establishes long-range dependencies but also demonstrates linear complexity related to input size. Inspired by Mamba, Vision Mamba \cite{vim} and Visual Mamba \cite{liu2024vmamba} were the first to introduce Mamba into the visual domain, achieving remarkable results. Additionally, numerous researchers have applied it to medical image segmentation, such as VM-UNet \cite{Ruan2024VMUNetVM}, U-Mamba \cite{Ma2024UMambaEL}, OCTAMamba \cite{zou2024octamamba}, and H-vmunet \cite{wu2024h}. However, challenges remain in skin lesion segmentation. For instance, the boundaries of the lesion areas are unclear, the sizes of the lesion regions vary, and there is significant information loss during downsampling. Additionally, we observed that directly applying Mamba to lesion segmentation results in reduced accuracy due to the inability to capture local fine-grained features.

To address the aforementioned issues, we propose SkinMamba, which combines the advantages of Mamba in learning global features and CNN in extracting local features. Specifically, the overall architecture follows a 5-level encoder-decoder structure. The core components of the model are the Scale Residual State Space Block (SRSSB) and the Frequency Boundary Guided Module (FBGM). In SRSSB, we introduce the Visual State Space Block (VSSB) to achieve efficient global modeling, while the Scale-Mixed Feed-Forward Layer (SMFFL) is used to extract cross-scale features, promoting the interaction of multi-level information and enabling multi-scale, multi-level feature representation in a global context. To mitigate the loss of boundary information caused by downsampling, we introduce the Frequency Boundary Guided Module (FBGM), which captures boundary cues from a frequency perspective and guides the decoder during the decoding process. Through this design, we developed the SkinMamba model, which maintains computational efficiency while effectively capturing both short- and long-range dependencies. It also addresses challenges such as unclear boundaries and varying lesion sizes in lesion images, further exploring Mamba's application in skin lesion segmentation.

The main contributions of our work are as follows:
\begin{itemize}
\item[$\bullet$]We proposed SkinMamba, a hybrid architecture based on Mamba and CNN, effectively combining the strengths of both. By addressing challenges such as varying lesion sizes and blurred boundary information in lesion segmentation, we further optimized the model architecture and explored the application of Mamba in this task.  
\item[$\bullet$]We introduced the Scale Residual State Space Block (SRSSB) to facilitate expert knowledge exchange across different levels in a global state and incorporated the Frequency Boundary Guided Module (FBGM) to achieve high-frequency restoration and boundary prior guidance.  
\item[$\bullet$]Extensive experiments on two open-source lesion segmentation datasets demonstrated that SkinMamba outperformed state-of-the-art methods in terms of mIoU, DSC, Acc, Spe, and Sen.
\end{itemize}

\section{Related Work}
\label{sec:related}

\subsection{Medical Image Segmentation}
FCN \cite{long2015fully} uses fully convolutional layers to extract image features, becoming a pioneer in image segmentation. However, medical image segmentation requires more detailed lesion feature learning compared to natural scene segmentation. Scale variation is a critical issue in medical image segmentation. UNet \cite{ronneberger2015u} addressed this by proposing an encoder-decoder architecture that effectively integrates low-level and high-level features, making it a cornerstone in medical image segmentation.

Since the introduction of UNet, many researchers have proposed variants to improve upon it. UNet++ \cite{zhou2019unetplusplus} replaces the traditional cropping and concatenation operations with dense convolutions, achieving tighter feature fusion in skip connections and reducing the impact of information loss during sampling. Attention-UNet \cite{oktay2018attention} integrates attention mechanisms into the UNet framework, focusing more on target regions and suppressing irrelevant information. To reduce information loss and improve overall performance, Res-UNet \cite{8589312} introduces Res-blocks to enhance feature transfer stability and training of deep networks. Dense-UNet \cite{8697107} utilizes Dense-blocks for efficient feature reuse, improving multi-scale information capture. U-Net v2 \cite{peng2023u} employs an innovative skip connection mechanism to mix high-level and low-level features, achieving refined feature fusion and enhancing the integration of features across different scales.

However, models using only CNNs suffer from reduced performance due to their inability to capture global features. To address this, researchers introduced Transformers from natural language processing (NLP) to computer vision \cite{vaswani2017attention}, developing Vision Transformer (ViT) for efficient global modeling. In medical image segmentation, TransUNet \cite{chen2024transunet} was the first to integrate ViT's feature learning capabilities into the U-Net encoder. By combining ViT's self-attention mechanism with U-Net's structural advantages, TransUNet improved feature representation accuracy and the model's understanding of complex image content. Swin-UNet \cite{swinunet} incorporates Swin Transformer network modules into the U-Net structure, enhancing feature learning capabilities. Attention Swin U-Net \cite{aghdam2023attention} operates with a cascaded attention mechanism and optimizes Swin-UNet. This cascading attention mechanism allows the model to more precisely capture and fuse important information during feature processing. 
% TransNorm \cite{Wang19TransNorm} introduces an attention mechanism with adaptive calibration of features, dynamically adjusting feature weights to achieve more precise feature fusion and enhance model representation. 
Although Transformers excel in global modeling, their attention mechanism faces high quadratic complexity in long sequence modeling, resulting in significant computational burden \cite{vaswani2017attention}. This is a considerable issue, especially for dense prediction tasks such as medical image segmentation.

\subsection{State Space Models (SSMs)}
In recent years, State Space Models (SSMs) \cite{Gu2021CombiningRC,kalman1960new} have gained attention in deep learning, particularly in dynamic system modeling. These models originated from state-space theory in control systems and have shown potential in capturing complex temporal dependencies. To address the issue of long-range dependencies that traditional temporal models struggle with, researchers have improved SSMs. For instance, S4 \cite{gu2022efficiently} uses a normalized diagonal matrix structure to reduce computational complexity, making it an effective alternative to CNNs and Transformers. Subsequent work, such as S5 \cite{smith2023simplified}, introduced a MIMO structure and optimized the performance of S4 layers through parallel scanning, proposing a new S5 layer. The Gated State Space layer further enhances model expressiveness by incorporating additional gating units in the S4 layer \cite{Mehta2022LongRL}. Recently, Mamba \cite{liu2024vmamba} has significantly improved the processing capability of SSMs for long-sequence tasks by introducing input-based dynamic parameter adjustments and utilizing highly optimized hardware algorithms. This advancement not only simplifies the model structure but also effectively addresses efficiency issues in processing long-sequence data, particularly in fields such as language and genomics. 

Following the success of Mamba, researchers have studied this framework in various fields, including image classification \cite{yang2024cmvim,GE2024128104,Ma2024UMambaEL,zou2024microscopic}, low-level \cite{zheng2024u,guo2024mambair,shi2024vmambair,chen2024mambauie}, point cloud analysis \cite{liang2024pointmamba,zhang2024point}, and remote sensing images \cite{chen2024changemamba,zhu2024samba,zhao2024rs}, making it a strong competitor to both CNNs and Transformers. In recent work, Mamba-based vision backbones have been applied to medical image segmentation. VM-UNet \cite{Ruan2024VMUNetVM} introduced the Visual State Space (VSS) module as a foundational block to build an asymmetric encoder-decoder structure, capturing extensive contextual information. U-Mamba \cite{Ma2024UMambaEL} and SegMamba \cite{xing2024segmamba} incorporated the SSM-Conv hybrid module, effectively leveraging the strengths of CNNs in local feature extraction and Mamba in long-range modeling. T-Mamba \cite{hao2024t} integrated shared positional encoding and frequency-based feature fusion into Vision Mamba to address limitations in frequency-domain spatial retention and feature enhancement. 

In this paper, we will further explore the application of the Visual State Space Model (VSSM) in skin lesion segmentation, thoroughly examining the limitations of Mamba in this task and addressing the remaining challenges in skin lesion segmentation, to optimize the model architecture accordingly.

\begin{figure}[tb]
  \centering
  \includegraphics[width=\linewidth]{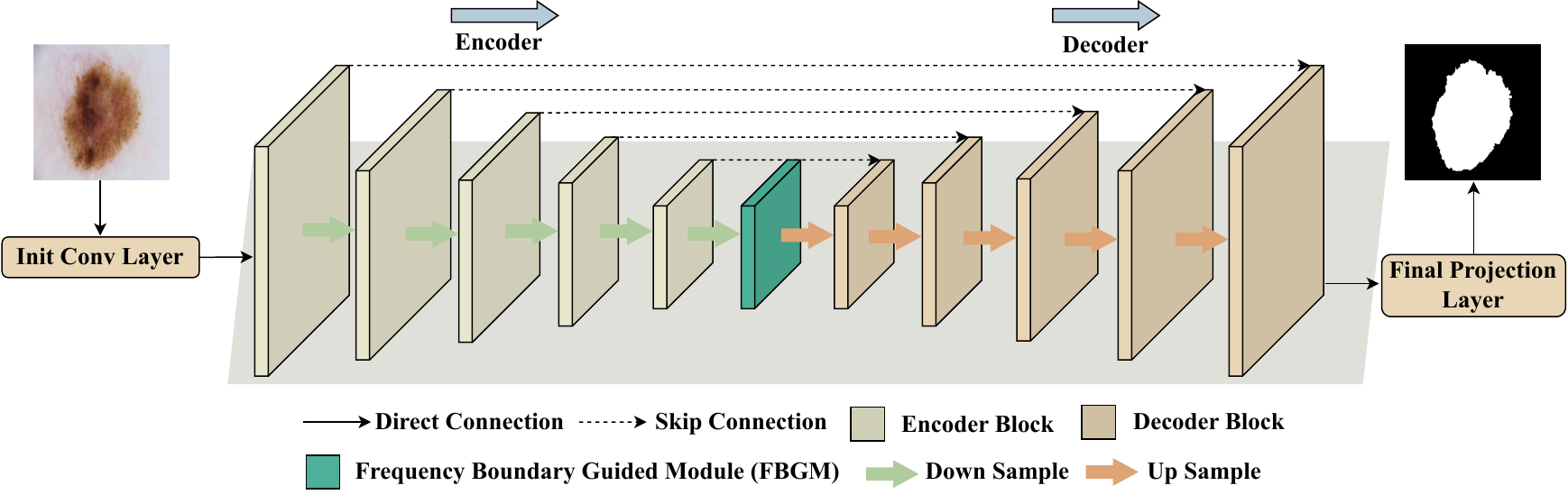}
  \caption{The illustration of SkinMamba architecture.}
  \label{fig:overall}
\end{figure}
\section{Method}
In this section, we first describe the entire pipeline, followed by details of the Encoder and Decoder Blocks. Finally, we elaborate on the two core components: the Scale Residual State Space Block (SRSSB) and the Frequency Boundary Guided Module (FBGM).

\subsection{Architecture Overview}
Our proposed SkinMamba is shown in \cref{fig:overall}. It includes an Init Conv Layer, an Encoder, an FBGM, a Decoder, a Final Projection Layer, and skip connections. Given an input image \(x \in \mathbb{R}^{H \times W \times 3}\), the Init Conv Layer first maps the channels of \(x\) to \(C\), with \(C\) defaulting to 16. Then, the Encoder performs deep feature extraction. Specifically, the Encoder consists of five stages, each comprising an Encoder Block and a downsampling layer. After passing through each stage, the input is mapped to a deeper feature space, while the image width and height are halved, and the number of channels is doubled. Similarly, the Decoder consists of five stages. Each stage includes a Decoder Block and an upsampling layer, where the image width and height are doubled, and the number of channels is halved. Notably, between the Encoder and Decoder, we introduce the FBGM to extract precise boundary cues from a frequency perspective, providing strong constraints for skin lesion segmentation and guiding image decoding. After the Decoder, the Final Projection Layer restores the channel count to match the segmentation target. For skip connections, we use simple addition operations to avoid unnecessary parameters, making it easier to validate the effectiveness of our architecture.

\begin{figure}[tb]
  \centering
  \includegraphics[width=\linewidth]{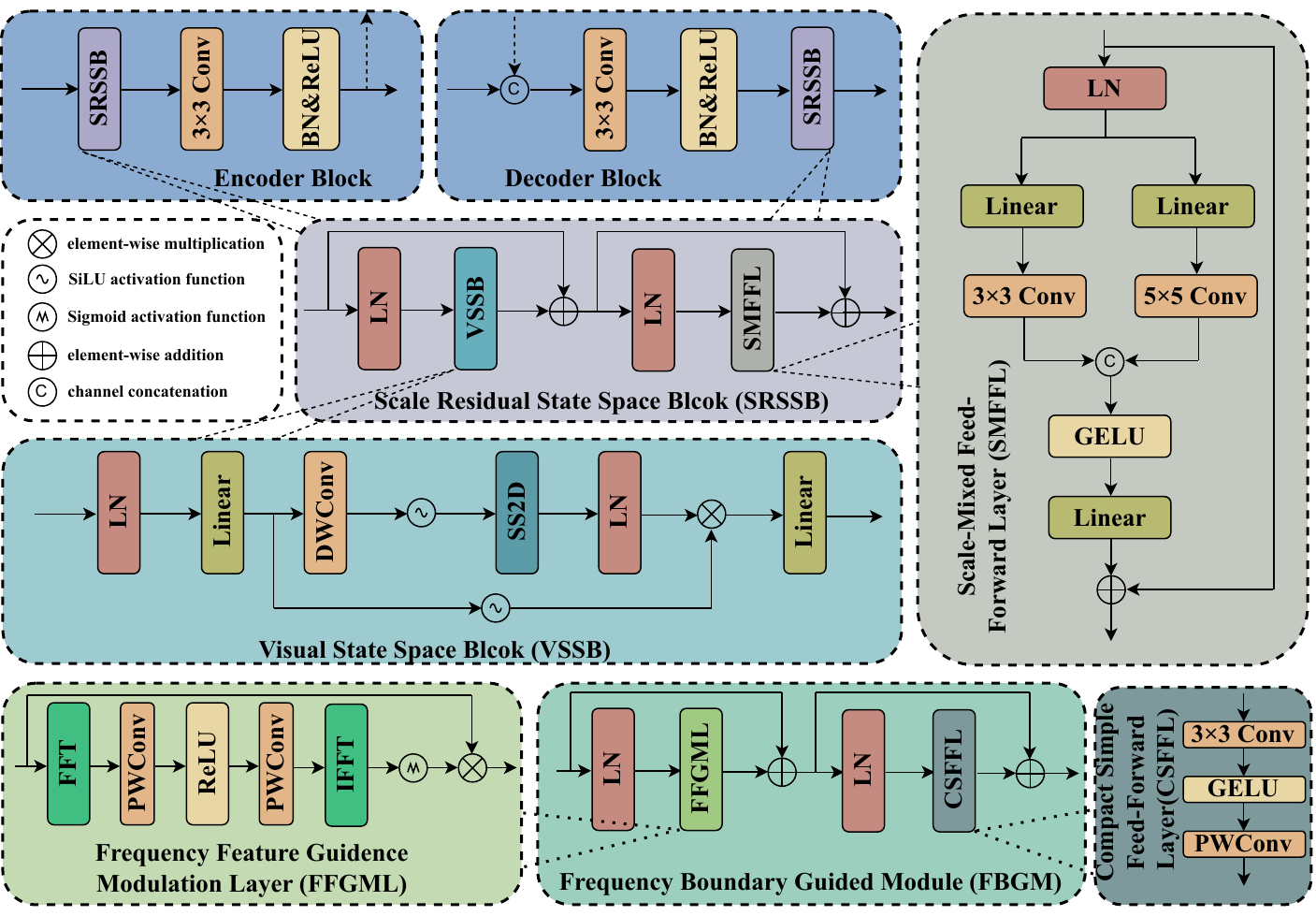}
  \caption{Key Components of SkinMamba.}
  \label{fig:detail}
\end{figure}

\subsection{Encoder and Decoder Blcoks}
Fusion and balance are the central themes of our architecture. Both the Encoder and Decoder blocks consist of CNN and SRSSB, combining Mamba's strength in learning global features with CNN's ability to extract local features.

As shown in \cref{fig:detail}, in the Encoder block, the input features first pass through the SRSSB, followed by a 3$\times $3 convolution, batch normalization (BN) layer, and ReLU activation function, producing the output features that are also used for skip connections. In the Decoder block, each Decoder block receives features from the FBGM or the previous Decoder block, along with the corresponding output features from the Encoder block's skip connections. These two sets of features are then aggregated along the channel dimension to form the fused features. The fused features pass through a 3$\times $3 convolution, BN layer, and ReLU activation function, and finally, the features are fed into the SRSSB.

\subsection{Scale Residual State Space Blcok}
The SRSSB is used to extract cross-scale mixed information from the global stream and to model long-range dependencies. As shown in \cref{fig:detail}, the core components of SRSSB are the Visual State Space Block (VSSB) and the Scale-Mixed Feed-Forward Layer (SMFFL). Specifically, it is divided into two stages. In the first stage, the input features are first passed through a Layer Normalization (LN) layer, followed by the VSSB to model global information. The second stage takes the features from the first stage, applies layer normalization to prevent mode collapse, and then processes them through the SMFFL to extract multi-scale features. This results in scale features under a global context, while residual connections help maintain consistent information flow across the SRSSB blocks, accelerating model convergence. The above process is defined as follows:
\begin{align}
X^{'}_{l}=X_{l-1}+{VSSB}(LN(X_{l-1}))
\\
X_{l}=X_{l}^{'}+{SMFFL}(LN(X_{l}^{'}))
\end{align}

Here, \( \text{LN}( \cdot ) \) denotes the layer normalization operation, \( X_{l-1} \) represents the input features, while \( X_{l}^{'} \) and \( X_{l} \) are the output features from the VSSB and SMFFL, respectively.

\subsubsection{Visual State Space Block.}
\cref{fig:detail} illustrates the VSSB. The input features \( X \) undergo layer normalization and a linear layer, generating two information streams. This process can be represented as:
\begin{align}
    F_{1}=F_{2}=Linear(LN(X))
\end{align}

Where \( \text{Linear}( \cdot ) \) represents processing through a linear layer, \( F_1 \) and \( F_2 \) are the inputs for the first and second information streams, respectively. In the first stream, the input features \( F_1 \) are processed by depth-wise separable convolution (DWConv) and the SiLU activation function, where preliminary features are extracted. These features are then refined through the 2D selective scanning (SS2D) module and normalized to produce the refined features \( F_1' \). In the second stream, the input features \( F_2 \) are processed by the SiLU activation function, where supplementary features \( F_2' \) are generated. The refined and supplementary features are then element-wise multiplied, merged, and passed through a linear layer to produce the final output of the VSSB, achieving comprehensive feature representation through detailed extraction and efficient global modeling. The above process is defined as follows:
\begin{gather}
    F_{1}^{'} = LN(SS2D(SiLU(DW(F_{1}))))
\\
F_{2}^{'} = SiLU(F_{2})
\\
F_{VSSB}=Linear(F_{1}^{'}\otimes F_{2}^{'})
\end{gather}

Where \( \text{DW}( \cdot ) \) denotes depth-wise separable convolution, \( \text{SiLU}( \cdot ) \) represents the SiLU activation function, \( \text{SS2D}( \cdot ) \) indicates 2D selective scanning, \( \otimes \) signifies element-wise multiplication, and \( F_{VSSB} \) denotes the final output features of the VSSB. 
\subsubsection{Scale-Mixed Feed-Forward Layer.}
As shown in \cref{fig:detail}, the SMFFL is a dual-branch multi-scale architecture. Specifically, the input features \( Map \) undergo layer normalization and are then divided into two branches. Each branch consists of a linear layer and a 3$\times $3 or 5$\times $5 convolution. First, the features are mapped to a low-dimensional space through the linear layer, and then different sizes of convolutions are used to capture features from the low-dimensional hidden layer. Subsequently, the low-dimensional features at different scales are aggregated along the channel to obtain scale-mixed features. Finally, GELU is applied for non-linear transformation, and a linear layer is used to remap the features back to the original high dimension, while the residual features are added to the scale-mixed features to facilitate gradient flow and propagation. The entire computational process of the module is defined as follows:
\begin{gather}
    Map^{'}=LN(Map)
    \\
    Map_{1}=f^{3\times 3}(Linear(Map^{'}))
    \\
    Map_{2}=f^{5 \times 5}(Linear(Map^{'}))
    \\
    Map_{residual}=Map
\\
    Map_{SMFFL}=Linear(GELU([Map_{1},Map_{2}]))\oplus Map_{residual}
\end{gather}

Where \( f^{x \times y}( \cdot ) \) represents the \( X \times Y \) standard convolution operation, \( [ \cdot , \cdot ] \) denotes the concatenation operation, \( \text{GELU}( \cdot ) \) represents the GELU activation function, and \( \oplus \) signifies element-wise addition. Additionally, we can observe that while the VSSB does not require positional encoding, the SMFFL plays a crucial role in promoting the interaction of multi-scale information flows. The SMFFL enhances synchronized operations, enabling each dimension and scale to contribute its specialized knowledge, thereby achieving precise and complementary detail extraction.

\subsection{Frequency Boundary Guided Module}
Generally, low-frequency information contains positional relationships and spatial information, while high-frequency information primarily includes specific boundary details. However, most methods reduce image resolution during encoding through downsampling to alleviate computational load. In such cases, decoding relies solely on the low-resolution image, leading to performance degradation and loss of high-frequency information. High-frequency information is crucial for the accurate segmentation of lesion areas on the skin. To address this, we developed the FBGM from a frequency perspective to provide sufficient boundary priors, accurately extract boundary clues, and guide the decoding process. As shown in \cref{fig:detail}, the Frequency Feature Guidance Modulation Layer (FFGML) and the Compact Simple Feed-Forward Layer (CSFFL) are the core components of FBGM. Given the input features \( K_{l-1} \), the proposed FBGM can be expressed as follows:
\begin{gather}
    K_{l}^{'}=K_{l-1}+FFGML(LN(K_{l-1}))
    \\
    K_{l}=K_{l}^{'}+CSFFL(LN(K_{l}^{'}))
\end{gather}

Where \( K_{l}' \) and \( K_{l} \) represent the output features of the FFGML and CSFFL, respectively.
\subsubsection{Frequency Feature Guidence Modulation Layer.}
\cref{fig:detail}  illustrates the FFGML. Specifically, given the input features \( F_0 \), they are transformed into the frequency domain using Fast Fourier Transform (FFT). We then process these frequency domain features with two layers of pointwise convolution (PWConv), interspersed with a ReLU activation function. Finally, the processed features are converted back to the original spatial domain using Inverse FFT, followed by a secondary modulation with the Sigmoid activation function. This modulated output interacts with the input features to obtain the updated feature representation of the original latent layer's hidden space. The above process is defined as follows:
\begin{gather}
    F_{mod}=\vartheta ^{-1} (PW(ReLU(PW(\vartheta (F_{0})))))
    \\
    \hat{F}_{mod}=Sigmoid(F_{mod})
    \\
    F_{FFGML}=\hat{F}_{mod} \otimes F_{0}
\end{gather}

Where \( \vartheta( \cdot ) \) denotes FFT, \( \vartheta^{-1}( \cdot ) \) represents Inverse FFT, \( \text{PW}( \cdot ) \) denotes pointwise convolution, and \( \text{Sigmoid}( \cdot ) \) represents the Sigmoid activation function.
\subsubsection{Compact Simple Feed-Forward Layer.}
Inspired by the Vision Transformer (ViT) \cite{dosovitskiy2020vit}, we designed a Compact Simple Feed-Forward Layer (CSFFL) for nonlinear transformation and dimensionality reduction. Given the input features \( X \), we first apply a 3$\times$3 convolution to increase the dimensionality, enhancing both the number of channels and local feature representation. This is followed by the introduction of nonlinearity through the GELU activation function. Finally, a PWConv is used to restore the original number of channels. The entire module is defined as follows:
\begin{align}
    X^{'}=PW(GELU(f^{3 \times 3}(X)))
\end{align}

\begin{table}[htbp]
\centering
\caption{Comparative experimental results on the ISIC2017 dataset. The best results are highlighted in bold fonts. “ ↑ ”and “ ↓ ” indicate that larger or smaller is better.
}
\begin{tabular}{p{2.5cm} >{\centering\arraybackslash}p{1cm} >{\centering\arraybackslash}p{1.8cm} >{\centering\arraybackslash}p{1.6cm} >{\centering\arraybackslash}p{1.6cm} >{\centering\arraybackslash}p{1.6cm} >{\centering\arraybackslash}p{1.6cm}}
\toprule
\textbf{Model} & \textbf{Year} & \textbf{mIoU(\%)$\uparrow$} & \textbf{DSC(\%)$\uparrow$} & \textbf{Acc(\%)$\uparrow$} & \textbf{Spe(\%)$\uparrow$} & \textbf{Sen(\%)$\uparrow$} \\
\midrule
UNet\cite{ronneberger2015u}           & 2015          & 75.97         & 86.34        & 95.53        & 97.75        & 84.47        \\
R2UNet\cite{alom2018recurrent}         & 2018          & 73.43         & 84.68        & 95.08        & 97.86        & 81.25        \\
UNet++\cite{zhou2019unetplusplus}         & 2019          & 77.85         & 87.55        & 95.91        & 97.94        & 85.82        \\
R2AttUNet\cite{10.1155/2021/6625688}      & 2021          & 75.07         & 85.76        & 95.24        & 97.17        & 85.63        \\
SwinUnet\cite{aghdam2023attention}       & 2022          & 67.93         & 80.90        & 93.75        & 96.69        & 79.11        \\
MISSFormer\cite{9994763}     & 2022          & 75.84         & 86.26        & 95.62        & \pmb{98.34}& 82.09        \\
MALUNet\cite{ruan2022malunet}        & 2022          & 74.69         & 85.51        & 95.15        & 97.10        & 85.46        \\
H2Former\cite{10093768}       & 2023          & 76.27         & 86.54        & 95.58        & 97.72        & 84.90        \\
EGE-UNet\cite{ruan2023ege}& 2023          & 76.50         & 86.68        & 95.65        & 97.88        & 84.55        \\
MHorunet\cite{wu2024mhorunet}       & 2024          & 78.16         & 87.73        & 95.77        & 97.15        & 85.99        \\
VMUNet\cite{Ruan2024VMUNetVM}         & 2024          & 77.24         & 87.16        & 95.78        & 97.82        & 85.62        \\
VMUNet v2\cite{zhang2024vm}      & 2024          & 75.25         & 85.88        & 95.34        & 97.47        & 84.71        \\
H-vmunet\cite{wu2024h}      & 2024          & 78.18         & 87.75        & 95.82        & 97.12        & 85.72\\
ULVM-UNet\cite{wu2024ultralight} & 2024   & 78.13         & 87.72        & 95.78& 97.59& 83.61        \\
SkinMamba      & -             & \pmb{78.30}& \pmb{87.83}& \pmb{96.00}& 97.99& \pmb{86.14}\\
\bottomrule
\end{tabular}
\label{table:isic2017}
\end{table}

\section{Experiments}
\label{sec:exp}
\subsection{Datasets}
In this section, we conducted extensive experiments on two public lesion segmentation datasets: the International Skin Imaging Collaboration 2017 and 2018 challenge datasets (ISIC2017 and ISIC2018), to train and evaluate the proposed model \cite{8363547,codella2019skin}. ISIC2017 and ISIC2018 contain 2,150 and 2,694 dermoscopic images with segmentation mask labels, respectively. Following previous studies \cite{ruan2022malunet,ruan2023ege}, we split the datasets into training and test sets at a 7:3 ratio. Specifically, the ISIC2017 training set consists of 1,500 images, while the test set contains 650 images. The ISIC2018 training set consists of 1,886 images, while the test set contains 808 images.

\begin{table}[tbp]
\centering
\caption{Comparative experimental results on the ISIC2018 dataset. The best results are highlighted in bold fonts. “ ↑ ”and “ ↓ ” indicate that larger or smaller is better.}
\begin{tabular}{p{2.5cm} >{\centering\arraybackslash}p{1cm} >{\centering\arraybackslash}p{1.8cm} >{\centering\arraybackslash}p{1.6cm} >{\centering\arraybackslash}p{1.6cm} >{\centering\arraybackslash}p{1.6cm} >{\centering\arraybackslash}p{1.6cm}}
\toprule
\textbf{Model} & \textbf{Year} & \textbf{mIoU(\%)$\uparrow$} & \textbf{DSC(\%)$\uparrow$} & \textbf{Acc(\%)$\uparrow$} & \textbf{Spe(\%)$\uparrow$} & \textbf{Sen(\%)$\uparrow$} \\
\midrule
UNet\cite{ronneberger2015u} & 2015          & 77.22         & 87.15        & 93.86        & 96.56        & 85.47        \\
R2UNet\cite{alom2018recurrent}         & 2018          & 71.74         & 83.55        & 92.36        & 96.41        & 79.74        \\
UNet++\cite{zhou2019unetplusplus}         & 2019          & 79.14         & 88.36        & 94.40        & 96.69        & 87.28        \\
R2AttUNet\cite{10.1155/2021/6625688}       & 2021          & 75.24         & 85.87        & 93.15        & 95.62        & 85.47        \\
SwinUnet\cite{aghdam2023attention}       & 2022          & 74.26         & 85.23        & 92.87        & 95.55        & 84.54        \\
MISSFormer\cite{9994763}     & 2022          & 77.94         & 87.60        & 94.11        & 96.89        & 85.48        \\
MALUNet\cite{ruan2022malunet}        & 2022          & 78.09         & 87.70        & 94.07        & 96.41        & 86.80        \\
H2Former\cite{10093768}       & 2023          & 77.33         & 87.21        & 93.89        & 96.57        & 85.56        \\
EGE-UNet\cite{ruan2023ege}        & 2023          & 78.90         & 88.20        & 94.25        & 96.17        & 88.29        \\
MHorunet\cite{wu2024mhorunet}       & 2024          & 79.40         & 88.52        & 94.47        & 96.70        & 87.55        \\
VMUNet\cite{Ruan2024VMUNetVM}         & 2024          & 74.14         & 85.15        & 93.03        & 96.54        & 82.10        \\
VMUNet v2\cite{zhang2024vm}      & 2024          & 78.25         & 87.80        & 94.09        & 96.25        & 87.38        \\
H-vmunet\cite{wu2024h}      & 2024          & 79.41         & 88.52        & 94.37        & 96.03        & 89.20        \\
ULVM-UNet\cite{wu2024ultralight} & 2024   & 78.74         & 88.10        & 94.29        & 96.68        & 86.85        \\
SkinMamba      & -             & \pmb{80.65}& \pmb{89.29}& \pmb{94.73}& \pmb{97.24}& \pmb{90.18}\\
\bottomrule
\end{tabular}
\label{table:isic2018}
\end{table}

\begin{table}[tbp]
    \centering
        \caption{Ablation study on VSSB.}\begin{tabular}{p{2cm} >{\centering\arraybackslash}p{1.8cm} >{\centering\arraybackslash}p{1.7cm} >{\centering\arraybackslash}p{1.7cm} >{\centering\arraybackslash}p{1.7cm} >{\centering\arraybackslash}p{1.7cm}}
        \toprule
\textbf{Setting} & \textbf{mIoU(\%)$\uparrow$} & \textbf{DSC(\%)$\uparrow$} & \textbf{Acc(\%)$\uparrow$} & \textbf{Spe(\%)$\uparrow$} & \textbf{Sen(\%)$\uparrow$} \\
        \midrule
        Ours    &      80.65&     89.29&     94.73&     97.24&     90.18\\
        Conv&      77.56&     87.36&     93.88&     96.12&     86.93\\
        self-attention&      78.25&     87.81&     94.06&     96.31&     87.62\\
        \bottomrule
    \end{tabular}

    \label{tab:vssb}
\end{table}

% \begin{table}[tbp]
%     \centering
%         \caption{Ablation study on SRSSB.}
%     \begin{tabular}{{p{1cm} >{\centering\arraybackslash}p{1.8cm} >{\centering\arraybackslash}p{1.7cm} >{\centering\arraybackslash}p{1.7cm} >{\centering\arraybackslash}p{1.7cm} >{\centering\arraybackslash}p{1.7cm}}}
%         \toprule
% \textbf{Ver.} & \textbf{VSSB} & \textbf{SMFFL} & \textbf{mIoU(\%)$\uparrow$} & \textbf{DSC(\%)$\uparrow$} & \textbf{Acc(\%)$\uparrow$} \\
%         \midrule
%         Ver 1 &      &       &      &     &     \\
%         Ver 2 &      &       &      &     &     \\
%         Ver 3 &      &       &      &     &     \\
%         Ver 4 &      &       &      &     &     \\
%         \bottomrule
%     \end{tabular}

%     \label{tab:srssb}
% \end{table}

\subsection{Implementation Details}
We implemented our SkinMamba using PyTorch 2.0.0 and trained it on an NVIDIA RTX 3090 with 24 GB of memory for 300 epochs with a batch size of 32. The input images are uniformly resized to 224 × 224. We employed data augmentation techniques such as random flipping and random rotation to prevent overfitting. We used the AdamW optimizer with an initial learning rate of \(1 \times 10^{-3}\), \( \beta_1 \) of 0.9, \( \beta_2 \) of 0.999, and weight decay of \(1 \times 10^{-4}\). Additionally, we applied a cosine annealing learning rate decay strategy and an early stopping mechanism. To ensure reproducibility, we set the random seed to 42.
\subsection{ Evaluation Metrics}
We used five metrics to evaluate the segmentation performance: Mean Intersection over Union (mIoU), Dice Similarity Score (DSC), Accuracy (Acc), Sensitivity (Sen), and Specificity (Spe). The mathematical formulations for these metrics are summarized as follows:
\begin{gather}
{mIoU}=\frac{{TP}}{{TP}+{FP}+{FN}}
   \\
    {DSC}=\frac{2{TP}}{2{TP}+{FP}+{FN}}
    \\
    {Acc=\frac{TP+TN}{TP+TN+FP+FN}}
    \\
    {Sen=\frac{TP}{TP+FN}}
    \\
    {Spe=\frac{TN}{TN+FP}}
\end{gather}

Where TP, FP, FN, TN represent true positive, false positive, false negative, and true negative. 

\subsection{ Comparison Results}
To validate the effectiveness of our approach, we compared SkinMamba with other state-of-the-art methods. Specifically, this comparison includes UNet \cite{ronneberger2015u}, R2UNet\cite{alom2018recurrent}, UNet++\cite{zhou2019unetplusplus}, R2AttUNet\cite{10.1155/2021/6625688}, SwinUnet\cite{aghdam2023attention}, MISSFormer\cite{9994763}, MALUNet\cite{ruan2022malunet}, H2Former\cite{10093768}, EGEUNet \cite{ruan2023ege}, MHorunet \cite{wu2024mhorunet}, VMUNet \cite{Ruan2024VMUNetVM}, VMUNet v2 \cite{zhang2024vm}, H-vmunet \cite{wu2024h}, UltraLight-VM-UNet \cite{wu2024ultralight}, and SkinMamba. \cref{table:isic2017} and \cref{table:isic2018} show the comparative results on the ISIC2017 and ISIC2018 datasets, respectively. Our proposed SkinMamba outperformed the other models in terms of mIoU, DSC, Acc, Spe, and Sen metrics.

% As shown in \cref{tab:vssb}, the visualized segmentation results demonstrate that SkinMamba achieved the closest results to the Ground Truth. It displayed greater sensitivity to lesion information and achieved better boundary delineation in segmentation outcomes.

% \begin{table}[tbp]
%     \centering
%         \caption{Ablation study on FBGM.}
%     \begin{tabular}{{p{1cm} >{\centering\arraybackslash}p{1.8cm} >{\centering\arraybackslash}p{1.7cm} >{\centering\arraybackslash}p{1.7cm} >{\centering\arraybackslash}p{1.7cm} >{\centering\arraybackslash}p{1.7cm}}}
%         \toprule
% \textbf{Ver.} & \textbf{FFGML}& \textbf{CSFFL}& \textbf{mIoU(\%)$\uparrow$} & \textbf{DSC(\%)$\uparrow$} & \textbf{Acc(\%)$\uparrow$} \\
%         \midrule
%         Ver 1 &      &       &      &     &     \\
%         Ver 2 &      &       &      &     &     \\
%         Ver 3 &      &       &      &     &     \\
%         Ver 4 &      &       &      &     &     \\
%         \bottomrule
%     \end{tabular}

%     \label{tab:fbgm}
% \end{table}

\begin{table}[tbp]
    \centering
        \caption{Quantitative comparisons with different combinations of the FBGM and SRSSB.}
    \begin{tabular}{{p{1cm} >{\centering\arraybackslash}p{1.8cm} >{\centering\arraybackslash}p{1.7cm} >{\centering\arraybackslash}p{1.7cm} >{\centering\arraybackslash}p{1.7cm} >{\centering\arraybackslash}p{1.7cm}}}
        \toprule
\textbf{Ver.} & \textbf{FBGM} & \textbf{SRSSB} & \textbf{mIoU(\%)$\uparrow$} & \textbf{DSC(\%)$\uparrow$} & \textbf{Acc(\%)$\uparrow$} \\
        \midrule
        Ver 1 &       &       &      77.90&     87.58&     93.84\\
        Ver 2 &       &       $\checkmark$&      79.90&     88.83&     94.60\\
        Ver 3 &      $\checkmark$ &       &      78.20&     87.77&     94.16\\
        Ver 4 &      $\checkmark$ &       $\checkmark$&      80.65&     89.29&     94.73\\
        \bottomrule
    \end{tabular}

    \label{tab:version_performance}
\end{table}

\subsection{ Ablation Experiment}
We conducted comprehensive ablation experiments on the ISIC2018 dataset to validate the effectiveness of each component in our proposed model. These experiments included evaluating the performance of VSSB, SRSSB, and FBGM components.
\subsubsection{Effects of Visual State Space Block.}We first analyzed the performance of CNN, Transformer, and Mamba within the framework. Specifically, we replaced the VSSB with 3$\times $3 convolutions from CNN and self-attention mechanisms from Transformer. As shown in \cref{tab:vssb}, replacing VSSB with convolutions or self-attention resulted in a decrease in performance.
% \subsubsection{Effects of Scale Residual State Space Blcok.} To evaluate the effectiveness of the components within SRSSB, \cref{tab:srssb} records the metrics of different models. We observe the following: (1) VSSB has the advantage of modeling global interactions from non-local regions, allowing it to focus on salient areas. (2) SMFFL effectively facilitates the exchange of expert knowledge across different levels, enabling the efficient segmentation of lesions of varying sizes.
% \subsubsection{Effects of Frequency Boundary Guided Module.}We explored the effects of the two components in FBGM, and the results are shown in \cref{tab:fbgm}. Both FFGML and CSFFL contribute from different perspectives, and the performance reaches its best when both components are used together.
\subsubsection{Effects of Key Components.}To further demonstrate the effectiveness of key components in SkinMamba, we performed ablation experiments by progressively removing or replacing each component. These components include SRSSB and FBGM. As shown in \cref{tab:version_performance}, each component significantly contributed to the model's performance. Specifically, the introduction of FBGM improved boundary detection and provided strong guidance for the decoder, while SRSSB enhanced cross-scale feature representation in a global context. Removing these components led to a decrease in both model accuracy and segmentation quality.

\section{Conclusions and Future Works}
\label{sec:con}
This study presents SkinMamba, an innovative skin lesion segmentation model that combines State Space Models (SSM) with Convolutional Neural Networks (CNN). By introducing the Scale Residual State Space Block (SRSSB) and the Frequency Boundary Guided Module (FBGM), SkinMamba effectively addresses the limitations of existing models in handling multi-scale lesion areas, boundary blurring, and Inconspicuous lesion regions. Extensive experiments on the ISIC2017 and ISIC2018 datasets demonstrate that SkinMamba outperforms current state-of-the-art methods in metrics such as mIoU, DSC, Acc, Spe, and Sen, showcasing its superior capability in accurately segmenting skin lesion areas. The success of SkinMamba is attributed to its complementary strengths in global and local feature extraction. SRSSB enables efficient global modeling, capturing cross-scale information, while FBGM enhances boundary detection by extracting boundary cues from the frequency domain.

Although SkinMamba has performed excellently in experiments, there is room for further optimization. Future research will focus on improving the model's adaptability and generalization, particularly in other medical image segmentation tasks. Additionally, we aim to reduce the model's complexity, ensuring that SkinMamba exhibits higher efficiency and adaptability on edge devices and in resource-constrained environments, thus expanding its potential application in real-time medical diagnosis systems.

\bibliographystyle{splncs04}
\bibliography{main}
\end{document}